\documentclass{jps-cp}
\usepackage{txfonts}
\usepackage{soul}
\PassOptionsToPackage{pdftex}{graphicx}
\usepackage{graphicx}
\usepackage{xcolor}
\usepackage{bm}
\usepackage{tikz}
\usepackage[hyphens]{url}
\usepackage{xurl} 
\usetikzlibrary{arrows.meta, positioning}

\title{Anomaly prediction in XRP price with topological features}

\author{Illia \textsc{Donhauzer}$^{1,3}$, Pierluigi \textsc{Cesana}$^{1}$, Tomoyuki \textsc{Shirai}$^{1}$
and ,Yuichi \textsc{Ikeda}$^{2}$}
\inst{$^{1}$Institute of Mathematics for Industry, Kyushu University, Japan \\
$^{2}$Graduate School of Advanced Integrated Studies in Human Survivability, Kyoto University, Japan\\
$^{3}$ La Trobe University, Australia}

\email{donhauzer.illia.501@m.kyushu-u.ac.jp, cesana.pierluigi.455@m.kyushu-u.ac.jp, shirai@imi.kyushu-u.ac.jp, ikeda.yuichi.2w@kyoto-u.ac.jp}

\recdate{March 10, 2026}

\abst{The aim of this research is to study XRP cryptoasset price dynamics, with a particular focus on forecasting atypical price movements. Recent studies suggest that topological properties of transaction graphs are highly informative for understanding cryptocurrency price behavior. In this work, we show that specific topological properties of the XRP transaction graphs provide important information about extreme XRP price surges, and can be used for more competitive prediction of~ anomalous price dynamics. 
} 

\kword{cryptocurrency, anomaly prediction, topological data analysis, LSTM, blockchain}

\begin{document}
\maketitle

\section{Introduction}
Due to increasing interest in the cryptocurrency market, the analysis and modeling of cryptoassets price dynamics has been a hot topic of research in recent years \cite{Aoyama, ikeda, Sankaewtong, Shirai1, Shirai2}. 
Reliable modeling allows efficient investment strategies to be devised, as well as contribute to the market's transparency and foster the broader adoption of blockchain-based technologies.
In our research, we model XRP price dynamics with special emphasis on price surges. 
Introduced by Ripple Labs in 2012, XRP is the native digital asset of the XRP ledger. It has established itself as one of the major cryptoassets by market capitalization and trading volume.

In the literature on cryptocurrency price modeling, the market trend is often forecasted using only historical prices and other financial and market indicators, without accounting for the financial network structure effect \cite{Akcora}. Compared to widely adopted standard forecasting methods, in our approach, on top of using rather classical market indicators, we also exploit more advanced data analysis techniques to extract deeper insights from XRP transaction data recorded in the distributed ledger that cannot be retrieved with conventional analytic methods. 

Topological data analysis (TDA) is a field that has recently emerged from algebraic topology and computational geometry. It aims to provide methods for analyzing the complex topological and geometric structures underlying the data \cite{Edelsbrunner}. The transaction data recorded on XRP ledger can be viewed as a weighted directed graph (a transaction graph), where XRP wallets are represented as nodes and the edges correspond to the transactions, with weights given by the transaction amounts. This graph representation captures relations between wallets, and topological data analysis allows one to analyze its structure. 

Recently, it has been shown that the transaction graph's topology provides important information about the future prices of cryptoassets. The work \cite{Akcora} studied the local topology of the Bitcoin networks and introduced a novel concept of chainlets, or Bitcoin network subgraphs. It was shown that the introduction of chainlet-based features can significantly improve the forecasting accuracy of machine learning models for price dynamics even when other more conventional factors, such as the historical price and the number of transactions, are already in the model. In the following, \cite{Abay} introduced the concept of Betti derivatives that can capture the rate of changes that occur in the topology of transaction graphs and showed the utility of the Betti derivatives in modeling Bitcoin prices. The introduction of Betti derivatives to machine learning models was shown to bring a significant gain in modeling accuracy. In \cite{Chakraborty}, high-resolution transaction-level data from the XRP Ledger were used to construct a cross-correlation tensor based on network embedding techniques, linking the structural dynamics of the XRP transaction network to movements in the XRP price. In \cite{ikeda}, the authors provided a detailed analysis of the correlation tensor spectra method for XRP transaction networks.
 
In our analysis, we exploit tools such as persistent homologies  \cite{Edelsbrunner, Zomorodian} and Betti numbers  \cite{Carlsson}, which are easily available in the corresponding Python packages Ripser \cite{ripser} and Networkx \cite{networkx}.  To further analyze the structure of the transaction graph, we study the frequency of appearances of specific subgraphs, or motifs. These motifs can capture specific transaction patterns and, as we show, a change in a number of certain motifs may indicate approaching changes in the price of the cryptoasset.
We show that the topological properties of the transaction graph carry important information on the future dynamics of the cryptoasset price. However, the important novelty of our research is that we show that the topological properties of the transaction graph become increasingly important when predicting anomalous price surges and can significantly improve their forecast accuracy when incorporated into machine learning models.

Our contributions can be summarized as follows:
\begin{itemize}
\item  we relate topological properties of the transaction graph with anomalous surges of the XRP~price and show that the topology of the transaction graph contains important information about extreme price surges,
\item motivated by the Betti derivatives recently introduced in \cite{Abay}, we propose a similar concept of {\it Betti increments} to quantify how topology evolves across a sequence of graphs,
\item we show that machine learning models incorporating information about the evolution of transaction graph topology outperform models based solely on standard market indicators when predicting anomalous price surges.
\end{itemize}

\section{Preliminaries}

In this section, we introduce the main definitions and tools used throughout the paper.

Let $\zeta = \{V, E, W\}$ be a weighted {\it undirected} graph with vertex set $V$, edge set $E \subseteq \{ \{u,v\}:\ u,v\in V, \ u\neq v\}$, and weight function $W : E \to (0,\infty)$. Let $\varepsilon_1, \varepsilon_2, \ldots, \varepsilon_n$ be an increasing sequence of scale values. For each $\varepsilon_k$, we define a subgraph
\begin{equation}
\zeta^{(\varepsilon_k)} = \{V^{(\varepsilon_k)}, E^{(\varepsilon_k)}, W\},
\end{equation}
where
\begin{equation}
V^{(\varepsilon_k)}
=
\left\{ 
u \in V \;:\; \exists\ v \in V,\ \{u,v\} \in E \text{ and }  W(\{u,v\}) \le \varepsilon_k
\right\},
\quad
E^{(\varepsilon_k)}
=
\left\{
e \in E : W(e) \le \varepsilon_k
\right\}.
\end{equation}
This construction yields a nested sequence of subgraphs,
\begin{equation}
\zeta^{(\varepsilon_1)} \subseteq \zeta^{(\varepsilon_2)} \subseteq \cdots \subseteq \zeta^{(\varepsilon_n)},
\end{equation}
which forms a graph filtration of $\zeta$.

Let $VR^{(\varepsilon_k)}$ denote the Vietoris--Rips complex~\cite{Sheehy} associated with the graph $\zeta^{(\varepsilon_k)}$, defined by
\begin{equation}
VR^{(\varepsilon_k)}
=
\left\{
S \subseteq V^{(\varepsilon_k)} :
W(e) \le \varepsilon_k \ \text{for all edges } e = \{u,v\} \text{ with } u,v \in S
\right\}.
\end{equation}
The collection $\{VR^{(\varepsilon_1)},VR^{(\varepsilon_2)},...,VR^{(\varepsilon_n)}\}$ constitutes the Vietoris--Rips filtration.

For each $VR^{(\varepsilon_k)}$, $k = 1,\ldots,n$, there are associated homology groups $H_p(VR^{(\varepsilon_k)})$, $p = 0,1,\ldots$, defined as the quotient groups of the cycle groups by the boundary groups. The Betti numbers $\beta_p^{(\varepsilon_k)}$ are the ranks of the homology groups $H_p(VR^{(\varepsilon_k)})$. For small values of $p$, Betti numbers admit intuitive interpretations: $\beta_0^{(\varepsilon_k)}$ equals the number of connected components in $\zeta^{(\varepsilon_k)}$, while $\beta_1^{(\varepsilon_k)}$ and $\beta_2^{(\varepsilon_k)}$ equal the numbers of loops and voids, respectively.
Thus, for the graph $\zeta$, the Betti sequences
\begin{equation}
\boldsymbol{\beta}_p
=
\left(
\beta_p^{(\varepsilon_1)},
\beta_p^{(\varepsilon_2)},
\ldots,
\beta_p^{(\varepsilon_n)}
\right),
\quad
p = 0,1,\ldots,
\end{equation}
describe its topological properties across different scales.

Consider a sequence of weighted undirected graphs $\{\zeta_1, \zeta_2,..., \zeta_T\}$, $\zeta_t = \{V_t, E_t, W_t\}, \ t=1,2,...,T.$ To quantify temporal changes in the topology of graphs $\zeta_t$, we introduce \emph{Betti increments}. Let $\boldsymbol{\beta}_p(t)$, $p = 0,1,\ldots$, denote the Betti sequences associated with $\zeta_t$. Betti increments are defined as \begin{equation}
\Delta \boldsymbol{\beta}_p(t)
:=
\boldsymbol{\beta}_p(t+1)
-
\boldsymbol{\beta}_p(t),
\quad
p = 0,1,\ldots.
\end{equation}
Analogously, the left Betti increments are defined by
\begin{equation}
\Delta_{-} \boldsymbol{\beta}_p{(t)}
:=
\boldsymbol{\beta}_p(t)
-
\boldsymbol{\beta}_p(t-1),
\quad
p = 0,1,\ldots.
\end{equation}

For a sequence of {\it weighted directed} graphs $\{\zeta_1,\zeta_2,...,\zeta_T\}$, $\zeta_t=\{V_t, A_t, W_t\}, \ t=1,2,...,T,$ such that $V_t$ is the set of vertices, $ A_t \subseteq \{ (u,v) \in V_t\times V_t, \ u\neq v\}$ is the  arc set, and $W_t : A_t \to (0,\infty),$ is the weight function, a widely used approach to quantify how the directed topology of $\zeta_t$ evolves over time is \emph{motif analysis}. Counting the frequencies of  certain subgraphs in $\zeta_t$, each  representing specific local topological features, one can   detect changes in the topology of the graphs $\zeta_t$ over time.

In the following section, we apply the topological data analysis tools presented here to gain insight into XRP price dynamics.

\section{Experiments}

\subsection{Data and methodology}

Our data are obtained from the official XRP data source~\cite{xrplink} (see also \cite{Chakraborty} for description of the data) and contain information on all XRP transactions from October 2017 to October 2021 (208 weeks in total). For each transaction, we observe the sender’s wallet ID, the receiver’s wallet ID, and the transaction amount. Transactions are partitioned into weekly intervals and, for each week, a transaction graph is constructed as follows. Let $\zeta_t = \{V_t, A_t, W_t\}, \ t=1,2,...,T,$ where $V_t$ is the set of wallet IDs that sent or received at least one transaction during week  $t$, $A_t \subseteq \{ (u,v) \in V_t\times V_t, \ u\neq v\}$ represents transactions between wallets, and $W_t : A_t  \to (0,\infty)$ assigns a transaction amount to each edge.

We consider the following forecasting task. Let $y_{t+1}$ denotes the XRP price increment in week $t+1$, defined as
\begin{equation}
    y_{t+1} = \text{price}_{t+1} - \text{price}_{t},
\end{equation} where $\text{price}_{t}$ and $\text{price}_{t+1}$ are the XRP prices at the end of weeks $t$ and $t+1$, respectively. The objective is to predict $y_{t+1}$ based on the market state in a previous week represented by the transaction graph $\zeta_t$ and a set of standard market indicators, such as XRP price at the end of the week, the weekly price increase and related variables.

To describe the topological properties of the graphs $\zeta_t$, we employ tools from topological data analysis. Since the computation of persistent homology and Betti numbers is computationally expensive for weighted directed graphs $\zeta_t$, we instead consider weighted undirected versions of the graphs $\widetilde{\zeta}_t$ defined as follows. $\widetilde{\zeta}_t=\{V_t, E_t, \widetilde{W}_t\},$ where 

\begin{equation}
E_t = \{\{u,v\}\subseteq V_t: \ (u,v) \in A_t \ {\rm or} \ (v,u) \in A_t \}, \quad \widetilde{W}(\{u,v\}) = W(u,v) + W(v,u).
\end{equation}

Then, for each week $t$ let us denote by 
\begin{equation} \boldsymbol{\beta}_p(t) = (
\beta^{(\varepsilon_{10})}_p(t),
\beta^{(\varepsilon_{20})}_p(t),
\ldots,
\beta^{(\varepsilon_{100})}_p(t)), \ p = 0,1,...,
\end{equation} Betti sequences associated with graphs $\widetilde{\zeta}_t,$ where $\varepsilon_{k},\ k = 10,20,...,100,$ are deciles of all transaction amounts made in week $t$. In our analysis, we focus on the case $p=0,1,$ as the calculation of higher-order Betti sequences ($p\geq2$) for large graphs requires substantial computational resources.

We quantify changes in the directed topology of the graphs $\zeta_t$ over time using Betti increments. Since at time $t$ only the current and past weeks transaction graphs are observable, we use left Betti increments
\begin{equation}\Delta_{-} \boldsymbol{\beta}_p(t)
=
\boldsymbol{\beta}_p(t)
-
\boldsymbol{\beta}_p(t-1),
\quad
p = 0,1,\end{equation} where the components of $\Delta_{-} \boldsymbol{\beta}_p(t)$ are given as $\Delta_{-}\beta^{(\varepsilon_{k})}_p(t)= \beta^{(\varepsilon_{k})}_p(t)-\beta^{^{(\varepsilon_{k})}}_p(t-1),$ $k = 10,20,...,100$.

We further analyze local topological properties of the graphs $\zeta_t$ using motif analysis. We count the frequencies of selected small subgraphs within each graph $\zeta_t$, which capture recurring local interaction patterns among wallets. In this analysis, we do not ignore directionality in the graphs $\zeta_t$. However, since calculating the frequencies of the motifs in the entire graphs $\zeta_t$ is computationally expensive, we filter each graph $\zeta_t$ retaining only the edges corresponding to transaction amounts in the top $1\%$ of all transaction amounts within a given week.

After exploring a range of candidate motifs, we identify those motifs whose temporal dynamics exhibit the strongest empirical association with XRP price movements. These motifs are illustrated in Figure~\ref{tab:cor_motif}. Table~\ref{tab:cor_motif} reports the correlations between the weekly changes in motif frequencies (the frequency in week $t$ minus the frequency in week $t-1$) and the change in XRP price $y_{t+1}$ in week $t+1$.

\begin{figure}
\caption{Motifs.}
\begin{tikzpicture}[
    >={Stealth[length=4mm,width=2.5mm]},
    node/.style={circle, draw=black, fill=cyan, minimum size=6pt, inner sep=0pt},
    edge/.style={->, gray, line width=1.6pt}
]
\hspace{1.15in}
\node[node] (a1) at (5,0.5) {};
\node[node] (b1) at (6,-0.5) {};
\node[node] (c1) at (7,-1.5) {};

\draw[edge] (a1) -- (b1);
\draw[edge] (b1) -- (a1);
\draw[edge] (b1) -- (c1);
\draw[edge] (c1) -- (b1);

\node at (6,-2) {motif 1};

\node[node] (a2) at (8.5,-1.5) {};
\node[node] (b2) at (9.7,0.5) {};
\node[node] (c2) at (10.9,-1.5) {};

\draw[edge] (a2) -- (b2);
\draw[edge] (b2) -- (a2);
\draw[edge] (b2) -- (c2);
\draw[edge] (c2) -- (a2);

\node at (9.8,-2) {motif 2};

\node[node] (a3) at (12,-1.5) {};
\node[node] (b3) at (13.2,0.5) {};
\node[node] (c3) at (14.4,-1.5) {};

\draw[edge] (a3) -- (b3);
\draw[edge] (b3) -- (a3);
\draw[edge] (b3) -- (c3);
\draw[edge] (c3) -- (b3);
\draw[edge] (c3) -- (a3);
\draw[edge] (a3) -- (c3);

\draw[edge,<->] (a3) -- (c3);

\node at (13.3,-2) {motif 3};

\label{tab:cor_motif} 

\end{tikzpicture}
\label{fig:cor_motif}
\end{figure}

\begin{table}[!htb]
\centering
\caption{Motif's pairwise correlations with the target variable.}
\begin{tabular}{lcccc}
\hline
\textbf{Variable} & \textbf{Pearson $r$} & \textbf{sign} & \textbf{Spearman $\rho$} & \textbf{sign}
\\
\hline
motif\_1\_inc                & 0.39 & yes & 0.16 & yes\\
motif\_2\_inc                & 0.20  & yes & 0.20 & yes\\
motif\_3\_inc                & 0.16  & yes & 0.14 & yes\\
\hline
\label{tab:cor_motif}
\end{tabular}
\end{table}

In addition to the topological features, we incorporate into our modeling a set of standard market indicators commonly used in cryptocurrency analysis. Furthermore, since cryptocurrency markets are strongly influenced by news, social media, and regulatory developments, we introduce a simple proxy for traders’ sentiment. 
We employ Google Trends data \cite{googletrends} to measure public attention by tracking the search frequencies of the terms “democrats” and “republicans.” 
Note that search frequencies for terms such as “XRP” or “cryptocurrency” typically increase after significant price movements and therefore provide limited predictive value. In contrast, searches related to political topics often rise prior to major political developments that may affect financial markets, making them potentially informative indicators.

We also include the Puell Multiple (data are available in \cite{puelllink} through a paid subscription), an important Bitcoin-based on-chain market metric that measures miners revenue relative to its long-term average. The Puell Multiple is defined as the ratio between the daily issuance value of Bitcoin (in USD) and its 365-day moving average. Values above the annual average indicate elevated current mining profitability, whereas lower values suggest that miners revenue is below its long-term mean~\cite{Akgül}.

Table~\ref{tab:description} summarizes the features calculated in a week $t$ that are used to predict the increment of XRP price $y_{t+1}$ in the following week. Since several components of $\Delta\boldsymbol{\beta}_p (t)$, $p=0,1$, are mutually correlated and some exhibit a weak association with $y_{t+1}$, we retain a single Betti-based feature,  $\Delta\beta^{(\varepsilon_{40})}_0(t)$, which shows the strongest association with $y_{t+1}$. In what follows, we denote this feature by $\Delta\beta_0$ for simplicity of exposition. After retraining our model multiple times, we found that models that incorporate $motif\_2\_inc$ exhibited the most stable performance compared to those that include other motifs. Therefore, we retain only $motif\_2\_inc$ in the final model. For the variable $trade\_volume\_1\%$, which represents the transaction volume of the top $1\%$ traders, we proceed as follows to identify these traders. For each wallet ID and each week, we compute the correlations between transaction volumes (the sum of all transaction amounts) in previous weeks and changes in the XRP price in the following weeks. Then, for each week, we rank the wallet IDs according to these correlations, and those in the top $1\%$ are identified as “top $1\%$ traders” for the given week.

\begin{table}[!htb]
\centering
\caption{Feature descriptions.}
\begin{tabular}{lc}
\hline
\textbf{Variable} & Description
\\
\hline
$price$ & XRP price at the end of a week $t$ \\
$price\_inc$ & XRP price increment in week $t$  (current week’s price minus the previous week’s price) \\
$trade\_volume$ & Sum of all transaction amounts \\
$trade\_volume\_1\%$ & Sum of all transaction amounts of top 1\% traders \\
$puell\_mult$       & Puell multiple at the end of a week\\
$puell\_mult\_inc$       & Puell multiple increment in a week\\
$sent\_inc$             & Sentiment of the market increment \\
$motif\_2\_inc$                & Frequency of motif 2 increment \\
$\Delta\beta_0$             & $\beta_0$ increment \\

\hline
\label{tab:description}
\end{tabular}
\end{table}

\begin{table}[!htb]
\centering
\caption{Pairwise correlations with the target variable.}
\begin{tabular}{lcccc}
\hline
\textbf{Variable} & \textbf{Pearson $r$} & \textbf{sign} & \textbf{Spearman $\rho$} & \textbf{sign}
\\
\hline
$price$                        & -0.20 & yes  & -0.24 & yes\\
$price\_inc$           & 0.21  & yes &  0.31 & yes\\
$trade\_volume$       & 0.14  & yes & -0.04  & no\\
$trade\_volume\_1\%$ & 0.26  & yes & 0.07  & no\\
$puell\_mult$       & 0.24 & yes &  0.13 & no\\
$puell\_mult\_inc$       & 0.23  & yes & 0.35 & yes\\
$sent\_inc$           & -0.18 & yes & -0.08 & no\\
$motif\_2\_inc$              & 0.20  & yes & 0.20 & yes\\
$\Delta\beta_0$            & 0.46 & yes & 0.28 & yes\\
\hline
\label{tab:cor}
\end{tabular}

\end{table}

Table~\ref{tab:cor} reports pairwise correlations between each feature in Table \ref{tab:description} and $y_{t+1}$. Among all features, the topological feature $\Delta\beta_0$ exhibits the strongest associations with $y_{t+1}$, suggesting that the topological properties of the transaction graphs $\zeta_t$ contain valuable information to predict the dynamics of the XRP price. In addition, we observe that the price increment exhibits significant autocorrelation at lag one week. This indicates that past price increments contain predictive information about future price movements. Therefore, the feature $\textit{price\_inc}$ is included in the model.

We train a long short-term memory (LSTM) neural network that integrates the features described in Table~\ref{tab:description} computed in week $t$ to predict $y_{t+1}$. Since the relationships between these features and $y_{t+1}$ may be nonlinear and involve temporal dependencies, we employ a multi-layer LSTM architecture, which is well suited to capturing such dynamics. For model training and evaluation, we adopt a chronological data split. Specifically, we use 60\% of the data (the first 130 weeks) for training, 20\% for validation using a walk-forward scheme over the subsequent 40 weeks, and the remaining 20\% for testing using a further 40-week walk-forward window. 

\begin{figure}
\centering
\includegraphics[height = 0.55\linewidth]{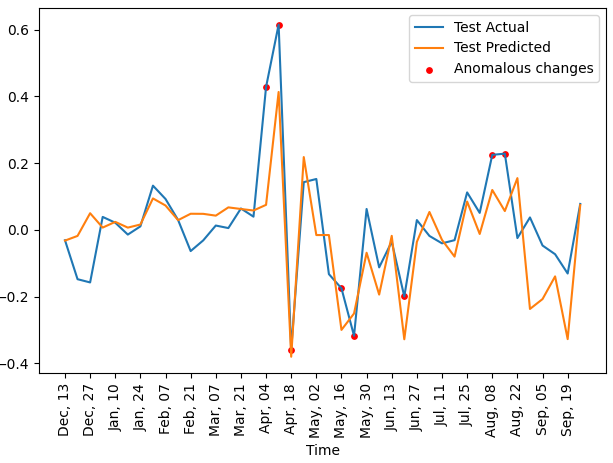}
\caption{XRP price increments in the test set (blue line) and their predictions (orange curve), red dots denote weeks with anomalous price increments.}
\label{pic:nnperf}
\end{figure}

The performance of the model on the test set is illustrated in Fig.~\ref{pic:nnperf}, where red dots indicate weeks with anomalous price changes. The pronounced price movement observed between April 4, 2021 and April 18, 2021, likely coincides with a favorable court decision involving XRP \cite{courtdecision}. Although information about this event was not explicitly included in the model, it anticipates both the price surge and the subsequent price correction.

\subsection{Contribution of the topological features}

To quantify the contribution of the topological features to the forecasting of XRP price anomalies, we employ two methodologies. The first is based on SHAP values, while the second relies on retraining the model with different subsets of features.

SHAP values provide a game-theoretic framework for explaining the outputs of machine learning models by quantifying the contribution of each feature \cite{Lundberg}. We select eight weeks from the test set (marked by red dots in Fig.~\ref{pic:nnperf}) during which the increase in XRP price was anomalous. For each of these weeks and for each feature in the model, we compute the corresponding SHAP values. We then find SHAP ranks of the features according to their average SHAP values over these eight weeks. To mitigate the effects of randomness associated with model training, we retrain the neural network $20$ times and find average SHAP ranks.

The first column of Table~\ref{tb:shap_anomaly} reports the average SHAP ranks computed for the eight anomalous weeks. We perform an analogous analysis throughout the whole test period, with the corresponding average SHAP ranks reported in the second column of Table~\ref{tb:shap_anomaly}. During weeks with atypical XRP price movements, the topological feature $\Delta\beta_0$ achieves the highest average SHAP rank, indicating its strong importance for predicting anomalous price changes. In contrast, during non-anomalous weeks, all features receive comparable SHAP ranks and $\Delta\beta_0$ becomes less influential. It demonstrates that the topological feature $\Delta\beta_0$ becomes increasingly important in predicting anomalous price movements.

\begin{table}[h!] \centering 
\caption{Average SHAP ranks of features.}
\begin{tabular}{lcc} \hline \textbf{Feature} & \textbf{Anomalous weeks} & \textbf{All weeks} \\ \hline 
$\Delta\beta_0$        & 2.7 & 4.6\\
$puell\_mult$            & 2.9 & 4.4\\
$price$                  & 3.0 & 4.5\\
$puell\_mult\_inc$       & 4.1 & 5.0\\
$sent\_inc$              & 5.3 & 5.3\\
$trade\_volume$          & 6.6 & 5.5\\
$motif\_2\_inc$          & 6.7 & 5.4\\
$trade\_volume\_1\%$     & 6.7 & 5.1\\
$price\_inc$             & 7.2 & 5.4\\
\hline
\end{tabular}
\label{tb:shap_anomaly} \end{table}

To further quantify the impact of incorporating topological features, we retrain the model 20 times using different sets of features and compare their predictive performance in terms of average RMSE. Table~\ref{tab:rmse} reports the accuracy gains in the prediction accuracy (reductions in RMSE) achieved by models with various combinations of features compared to $basic$ model, which refers to a model that excludes topological features ($\Delta\beta_0$ and $motif\_2\_inc$).

\begin{table}[h!] 
\centering 
\caption{RMSEs of models.}
\begin{tabular}{lcc} \hline \textbf{Features} &  \textbf{Anomalous weeks \% gain} &  \textbf{All weeks \% gain} \\ \hline 
$basic$     &    0\% & 0\% \\
$basic+\Delta\beta_0$             & 9.9\% & 4.6\% \\
$basic+motif\_2\_inc$           & 2.3\% &  0.7\% \\
$basic+\Delta\beta_0+motif\_2\_inc$    & 12.4\% & 5.5\% \\
\hline
\end{tabular}
\label{tab:rmse} \end{table}

The results confirm the conclusions drawn from the SHAP analysis and indicate that the topological properties of the transaction graphs carry important information about price changes. Specifically, topological features obtained from the transaction graphs become increasingly important when predicting anomalous XRP price movements.

\section{Conclusion}

We used topological data analysis tools to analyse the  transaction data recorded in the XRP ledger and construct novel topological features for modeling the XRP price dynamics. Our results demonstrated that the topological structure of the transaction graphs encodes important information about the dynamics of the XRP price, especially during periods of anomalous price behavior. We developed an LSTM-based model and, through SHAP analysis, showed that topological features contribute significantly to the forecasting of anomalous XRP price changes. We showed that incorporating these features into machine learning models consistently improves the accuracy of forecasting, leading to a reduction in error, especially during periods of extreme price movements.

\section{Acknowledgment}
This study was supported by Ripple Impact Fund 2022-247584 (5855). 
Pierluigi Cesana holds an honorary appointment at La Trobe University and is a member of GNAMPA.
Tomoyuki Shirai was supported in part by JSPS KAKENHI Grant Numbers JP22H05105  and JP23K25774.

\end{document}